\documentclass{PoS}

\title{The AGN phenomenon: open issues}

\ShortTitle{The AGN phenomenon}

\author{\speaker{Volker Beckmann}\\ 
        Fran\c{c}ois Arago Centre, APC, Universit\'e Paris Diderot, CNRS/IN2P3, CEA/Irfu, Observatoire de Paris, Sorbonne Paris Cit\'e, 13 rue Watt, 75013 Paris, France\\
        E-mail: \email{beckmann@apc.univ-paris7.fr}}

\author{Chris R. Shrader\\
        NASA Goddard Space Flight Center, mail code 661, Greenbelt, MD 20771, USA\\
        E-mail: \email{Chris.R.Shrader@nasa.gov}}

\abstract{The aim of this short paper is to motivate and encourage research in the field of Active Galactic Nuclei (AGN). Here we summarize the main open questions concerning the central engine. Is the central black hole rapidly spinning and can we prove this? What is the dominant accretion mechanism in AGN? Why do some AGN form jets while others don't and how do the jets originate? What keeps jets collimated out to distances of 100~kpc? Is the emission of blazars dominated rather by synchrotron self-Compton or by external Compton processes? Which parameters are important in the unified model? 
We outline the status of related research, formulate the questions and try to hint at research projects able to tackle these fundamental topics. Deep surveys, polarization measurements, improved models, faster and more accurate simulations as well as bridging the gap in the MeV range can be part of the tools to bring us closer to an understanding of AGN.}

\FullConference{"An INTEGRAL view of the high-energy sky (the first 10 years)"
9th INTEGRAL Workshop and celebration of the 10th anniversary of the launch,\\
		October 15-19, 2012\\
		Bibliotheque Nationale de France, Paris, France}

\begin{document}

\section{Introduction}

Active Galactic Nuclei (AGN) have been a rich field of research since their discovery in the 1940s. By now, we are quite certain about a number of characteristics of AGN. The central engine is believed to contain a super massive black hole in the range $10^4 - 10^{10} \rm \, M_\odot$, which converts the potential energy of matter in an accretion process to radiation and particle outflow. As far as the current status of research, there are no ``naked'' super massive black holes, rather all of them reside in galaxies. It seems that AGN activity at large has been higher in the past, with a maximum around redshift $z = 1 -3$, depending on source type and central engine size. We can observe AGN now back to redshifts of $z > 7$ and in all types of host galaxies. The general aspects of AGN that emerges from almost 60 years of research have been summarized and discussed in a recent text book \cite{AGNbook}. Here we want to raise the most prominent questions concerning the central engine, as we see them today.

\section{The Central Engine}

At the center of the AGN we commonly believe that a super massive black hole is accreting and thus providing the main powering mechanism. The putative angular momentum, or "spin", of this black hole has important implications on the overall behavior of the system. From a spinning black hole it is possible to extract energy  at the expense
of its rotational energy through the so-called Penrose process.
This involves
the spinning black hole's so called ergosphere; an ellipsoidal region
just outside its event horizon. Matter within the ergosphere
co-rotates with the black hole, but because that region is outside
the event horizon the matter can escape and carry energy drawn from
the black hole with it. 
The task here is to determine whether a large fraction of AGN host maximal spinning (Kerr) black holes, or whether most of them can be modeled assuming a non-rotating (Schwarzschild) case. 
The iron K$\alpha$ line at $E = 6.4 \rm \,
keV$ seems to be the best tool for assessing black hole rotation, because the line profile, if formed close enough to the event horizon, will be relativistically broadened and asymmetric for fast spinning black holes (e.g. \cite{Fabian00}). Some colleagues have raised questions though, how precisely one can measure the spin based on the line profiles, considering uncertainties in the measurement of the underlying continuum based on complex absorption (e.g. \cite{Miller08}) and intercalibration problems (e.g. \cite{Miyakawa09}), and also other models have been brought forward which can explain broadened lines. 

X-ray polarization experiments
may offer a powerful an independent tool, but unfortunately the NASA's GEMS mission was canceled in 2012. 
In order to accurately measure the iron line profile, the continuum has to be covered from significantly below to
significantly above $6.4 \rm \, keV$. Future X-ray
spectrometers may be able to exploit the iron L-shell lines, in the
photon-rich $\sim 1 \rm \, keV$ spectral region for this purpose as well.
It can be expected that {\it Astro-H} with its broad band and high spectral resolution will help to
settle the question about the spin of the central engine.

The accretion process of matter onto the super massive black hole raises another range of questions. 
Our standard model describes the accretion flow through an axisymmetric accretion disk, the so-called $\alpha$-disk. But in many cases we see AGN which lack the signature of an accretion disk in the UV (e.g. \cite{Nemmen10}). 
In such cases alternative
mechanisms such as the advection dominated accretion flow
(ADAF; e.g. \cite{Narayan95}) may be occurring. 
Is spherically symmetric accretion possible, at least in some cases?
This is a field that is likely to benefit from sophisticated
numerical simulations that will inevitably be facilitated by faster
computers and increasingly clever algorithms. These will have to
include magneto-hydrodynamic effects, and will ultimately need to
be done in 3 dimensions if we are to achieve a realistic picture.
Although at some level accretion disk structure may be directly
observable, for example as inferred from temperature gradients in
the innermost region of nearby Seyferts (e.g. NGC 3115 \cite{Wong11}) or in {\it HST} direct
imaging of M~87 (e.g. \cite{Perlman11}), this will help us only for a few of the nearest
AGN.

We also see hard X-ray emission  from non-beamed AGN. In this case we observe an inverse Compton component which needs a relativistic charged particles to be created. 
 Commonly this is modeled by an ambient hot cloud or disk
corona. The geometry of this electron
plasma cloud is unknown. A variety of ideas have been considered
such an arrangement of clouds, forming a "patchy corona" \cite{Haardt94}. This
scenario allows for transmission of the soft-excess components
sometimes seen, but is problematic for modeling observed
variability. How close to the black hole is the corona, how far out
does it extend? Spectroscopy alone cannot tell us, instead
time-resolved spectroscopy is our most promising tool. In recent
years, advances in this field have been driven by observations
using e.g. {\it Chandra},
{\it XMM-Newton} and {\it INTEGRAL}. To continue to make gains
though, parallel advances in modeling are needed that can be applied to
these data, as is happening already today for the
photon-rich Galactic X-ray binaries.

Historically, AGN have been divided into radio loud and radio quiet objects, i.e. those with and without a prominent jet emission. 
Still we do not know how these jets are launched, how they remain collimated over distances of 100s kpc, and what particles dominate them. 
From current
observations we stand at an unprecedented level in terms of the
empirical picture. From VLBI radio imaging, one is able to pin-point
the starting point of the jet and begin to see how
the jet evolves (e.g. \cite{Hada11}). We can directly observe spatial
components, or "blobs", as they are being emitted and then as they
travel downstream. Shocks and knots in the jet have been identified
and sensitive observations in other wavebands allow us to correlate
morphological evolution in the radio to the multi-wavelength light
curves (e.g. \cite{Marscher11}). But the main question still persists: how are the jets
formed in the first place? This question is tightly connected to the accretion process powering the AGN. 
One way to raise the jet can be to collimate the disk wind, but for a precise model the disk structure must be known. In addition, magnetic fields appear to be important in both, the launching and the collimation of the jet, but the exact geometry is unknown. As mentioned above, another source for powering the jet might be extraction of rotational energy from the Kerr black hole in the center. 

On top of these uncertainties lies the question why some AGN do develop jets while
some, apparently most, do not. The Galactic black hole binaries seem
to be most luminous when the jet is absent \cite{Fender04}, whereas the most
luminous quasars seem to sustain jet production. 
Once the matter is
launched and collimated into a jet, what is the acceleration
mechanism leading to the relativistic particles we observe, or more
accurately whose existence we infer from the observed spectral
energy distribution involving the synchrotron and inverse Compton
processes? It is perplexing that after many years of blazar
research, we are unclear as to what particles dominate the outflow.
Is this a hadronic or a leptonic jet? Do we have to consider
protons, which need more energy than electrons to be accelerated up
to the required speeds but which can more naturally explain
the remarkably collimated and expansive structures extending to
$\sim 100 \rm \, kpc$ distances? This begs the next open question:
what collimates the jets so effectively over such vast distances,
that can exceed for example the diameter of our Galaxy? Here
polarization mapping might give us more hints on the constraining
magnetic field, and clearly higher-resolution, higher-cadence radio
observations along with higher fidelity numerical modeling are our
pathway towards understanding jet structures.

The modeling of the spectral energy distribution is a powerful tool to gain insight into the physics underlying blazars. With the well sampled SEDs which have been compiled of a number of blazars, synchrotron emission appears to be the dominant process from the radio to the optical/UV/X-ray domain. But it is not clear at present where the seed photons for the second SED component come from. 
Are blazars predominantly one-zone synchrotron-self Compton emitters (SSC), where the synchrotron branch provides seed
photons for the IC process? Do we see instead a multi-component SSC?
Or is there an alternative seed-photon source such as the accretion
disk or  broad line region, giving rise to external Compton (EC)
processes? 

A challenge in this research is the fact that model parameters are only constrained if the SED is densely sampled in time and energy. 
In addition, only a limited number of blazars 
(Mrk~421, Mrk~501, 3C~454.3, 3C~279 ...) have been studied in sufficient depth, and those sources might represent special cases among the blazar class. 
In addition, most high-quality SEDs have been obtained during flaring state, which might not be representative for the average emission state of blazars.

{\it Fermi}/LAT has detected a number of of extragalactic gamma-ray
sources which are not bona-fide blazars, such as radio galaxies. This raises the question as to what drives them. Can we classify all
such sources being dominated by a mis-aligned jet or by cosmic rays
originating from star forming processes, a situation we find in
NGC~253 and M~82? How do the gamma-ray bright (and radio loud)
Narrow Line Seyfert 1 and radio galaxies fit this picture? The
ongoing {\it Fermi} survey is compiling a
database of cosmic gamma-ray sources unprecedented in quality and
quantity and offers our best opportunity to assess these classes. In the future, the Cherenkov Telescope Array (CTA; \cite{CTA}) is expected to provide insight into which of these sources produce significant TeV emission, with an expected number of 1000 source detections, most of them expected to be blazars. 

Although it would be of great interest to connect blazar or other
gamma-ray AGN SEDs in the X-ray band with sensitive observations in
the MeV range, telescopes to accomplish this do not currently exist. Closing this gap in which most of the FSRQ type blazars peak and in which one expects to find the transition from the thermal to the non-thermal inverse Compton emission will be crucial, and the scientific impact of an MeV telescope would indeed be large \cite{Lebrun10,Diehl10}. 

\begin{figure}
\begin{center}
\includegraphics[angle=270,width=.8\textwidth]{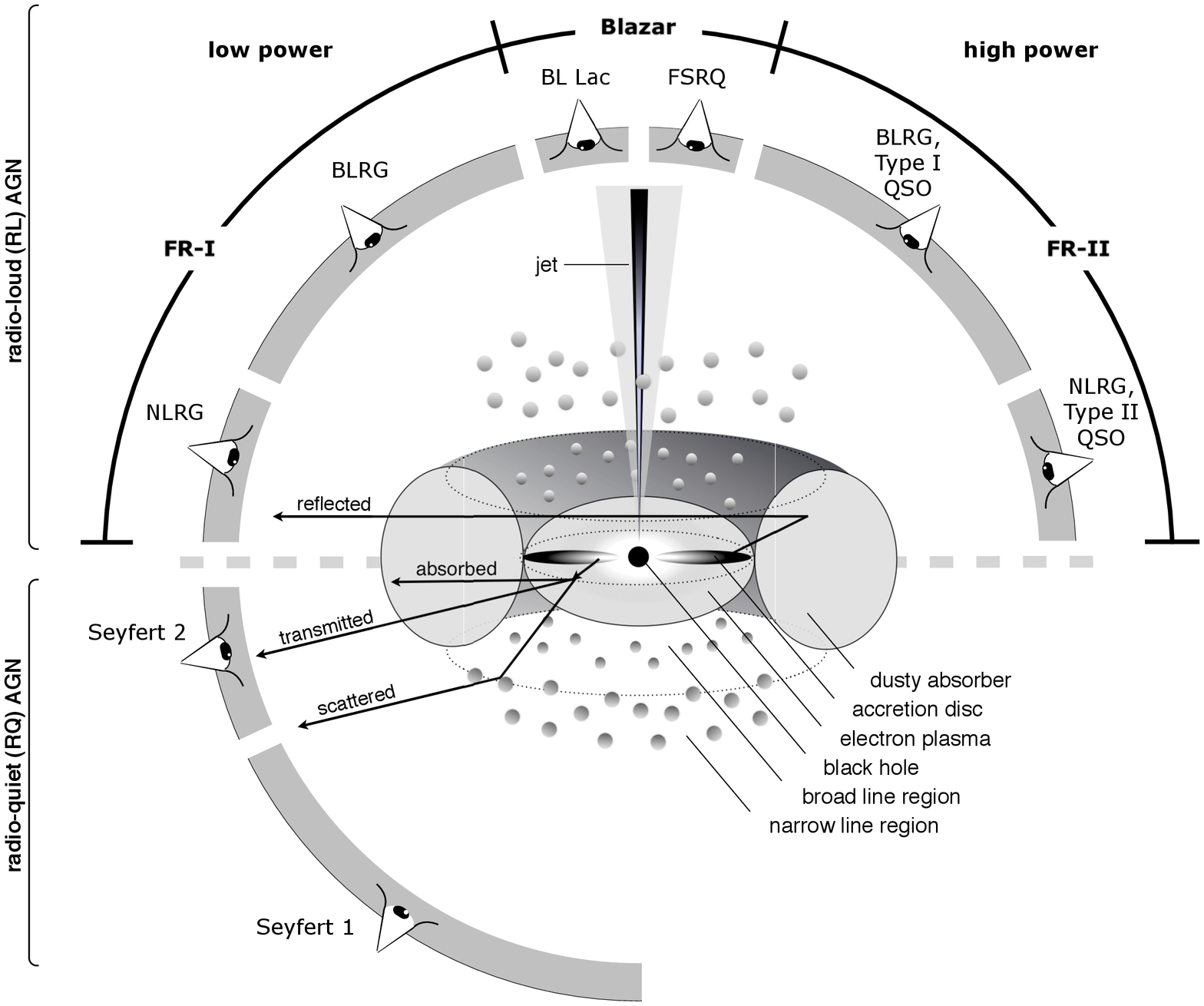}
\caption{Schematic representation of our
understanding of the AGN phenomenon in the unified scheme \cite{AGNbook}. The type
of object we see depends on the viewing angle, whether or not the
AGN produces a significant jet emission, and how powerful the
central engine is. Note that radio loud objects are generally
thought to display symmetric jet emission. Graphic courtesy of Marie-Luise Menzel (MPE).}
\label{fig:AGNgrafik}
\end{center}
\end{figure}

The unification scheme of AGN is generally accepted. The two main driving parameters are here the orientation of the AGN with respect to the line of sight, and the radio loudness. The first one determines the detectability of the central engine and the broad line region in the optical domain, the latter one indicates whether or not the AGN produces a significant jet.
Still, this simple unification scheme as it is summarized in
Fig.~\ref{fig:AGNgrafik} fails to explain all the different
varieties of AGN phenomena that we observe. 
Other important
considerations may be the mass of the central black hole, the
accretion rate, or the specific geometry of the absorber. Expanded
samples of Seyfert galaxies should help bolster statistics needed to
clearly identify the underlying factors. The SDSS has been
invaluable in this respect and future progress can be expected from
SDSS-III, {\it Gaia}, LSST, {\it Euclid}, and other surveys which will catalog
$\sim 10^7$ AGN including their redshifts. In the low-luminosity
regime, work remains in order to bridge the gap between the Galactic
black hole binaries and the super massive black holes. The
Ultra-luminous X-ray sources (ULX) are 
candidate examples of intermediate mass black holes (IMBH; \cite{Colbert99}) which
could help bridge the gap, but further study and in particular
improved classification of their non-X-ray counterparts will be necessary to
settle this question. Other LLAGN classes need to be separated
beyond ambiguity from the non-active galaxies. In particular H\,{\sc
ii} galaxies and LINER tend to become indistinguishable below some
signal-to-noise threshold \cite{Cid10}. The forthcoming large survey telescopes
surveys should bring clarification. Finally, the illusive link
between AGN and non-active super massive black holes, like Sgr~A* in our
very own galaxy, needs to be understood.

AGN research remains a rich field, worthy of our investments of
time, energies and talents that will continue to provide
 unexpected future insights into the nature of the Universe we live in.\\

{\it Acknowledgement}: We thank the anonymous referee for the constructive comments.


\begin{thebibliography}{99}
\bibitem{AGNbook}
Beckmann, V. \& Shrader, C. R. 2012, ``Active Galactic Nuclei'', 380 pages, Wiley-VCH
\bibitem{Fabian00}
Fabian, A.~C., Iwasawa, 
K., Reynolds, C.~S., \& Young, A.~J.\ 2000, PASP, 112, 1145
\bibitem{Miller08}
Miller, L., Turner, T. J., Reeves, J. N. 2008, A\&A, 483, 437
\bibitem{Miyakawa09}
Miyakawa, T., Ebisawa, K., Terashima, Y., et al.\ 2009, PASJ, 61, 1355 
\bibitem{Nemmen10} 
Nemmen, R.~S., Storchi-Bergmann, T., Eracleous, M., \& Yuan, F.\ 2010, IAU Symposium, 267, 313 
\bibitem{Narayan95}
Narayan, R., \& Yi, I.\ 1995, ApJ, 452, 710 
\bibitem{Wong11}
Wong, K.-W., Irwin, J.~A., Yukita, M., et al.\ 2011, ApJL, 736, L23 
\bibitem{Perlman11}
Perlman, E.~S., Adams, S.~C., Cara, M., et al.\ 2011, ApJ, 743, 119 
\bibitem{Haardt94}
Haardt, F., Maraschi, L., \& Ghisellini, G.\ 1994, ApJL, 432, L95 
\bibitem{Hada11}
Hada, K., Doi, A., Kino, M., et al.\ 2011, Nature, 477, 185 
\bibitem{Marscher11}
Marscher, A., Jorstad, S.~G., Larionov, V.~M., Aller, M.~F., L{\"a}hteenm{\"a}ki, A.\ 2011, Journal of Astrophysics and Astronomy, 32, 233 
\bibitem{Fender04}
Fender, R.~P., Belloni, T.~M., \& Gallo, E.\ 2004, MNRAS, 355, 1105 
\bibitem{CTA}
Actis, M., Agnetta, G., Aharonian, F., et al.\ 2011, Experimental Astronomy, 32, 193 
\bibitem{Lebrun10}
Lebrun, F., Aharonian, F., Beckmann, V., et al.\ 2010, $8^{\rm th}$ Integral Workshop ``The Restless Gamma-ray Universe'', PoS, 115, 34
\bibitem{Diehl10}
Diehl, R. 2010, $8^{\rm th}$ Integral Workshop ``The Restless Gamma-ray Universe'', PoS, 115, 35
\bibitem{Colbert99}
Colbert, E.~J.~M., \& Mushotzky, R.~F.\ 1999, ApJ, 519, 89
\bibitem{Cid10}
Cid Fernandes, R., Stasi{\'n}ska, G., Schlickmann, M.~S., et al.\ 2010, MNRAS, 403, 1036 
\end{thebibliography}
\end{document}